\documentstyle[12pt,preprint,psfig]{aastex}

\shorttitle{Star-Forming Lensed System: First Light at $z\simeq$5.6?}
\shortauthors{Ellis et al.}

\begin{document}

\title{A Faint Star-Forming System Viewed Through the Lensing Cluster 
Abell\,2218: First Light at $z\simeq$5.6?}
\author{Richard Ellis}
\affil{Astronomy Department, California Institute of Technology, Pasadena, CA 91125}
\email{rse@astro.caltech.edu} 
\author{Mike Santos}
\affil{Theoretical Astrophysics, California Institute of Technology, Pasadena, CA 91125}
\email{mrs@astro.caltech.edu}
\author{Jean-Paul Kneib}
\affil{Observatoire Midi-Pyrenees, 14 Ave Edouard Belin, F-31400 Toulouse, France}
\email{kneib@ast.obs-mip.fr}
\author{Konrad Kuijken}
\affil{Kapteyn Institute, PO Box 800, NL-9700 AV Groningen, Netherland}
\email{kuijken@astro.rug.nl}

\begin{abstract}

We discuss the physical nature of a remarkably faint pair of Lyman
$\alpha$-emitting images discovered close to the giant cD galaxy in the
lensing cluster Abell 2218 ($z$=0.18) during a systematic survey for
highly-magnified star-forming galaxies beyond $z$=5. A well-constrained
mass model suggests the pair arises via a gravitationally-lensed source
viewed at high magnification. Keck spectroscopy confirms the lensing
hypothesis and implies the unlensed source is a very faint ($I\sim$30)
compact ($<$150 $h_{65}^{-1}$ pc) and isolated object at $z$=5.576
whose optical emission is substantially contained within the Lyman
$\alpha$ emission line; no stellar continuum is detectable. The
available data suggest the source is a promising candidate for an
isolated $\sim$10$^6$ $M_{\odot}$ system seen producing its first
generation of stars close to the epoch of
reionization.\footnote{\footnotesize Using data obtained with the
Hubble Space Telescope operated by AURA for NASA and the W.M. Keck
Observatory on Mauna Kea, Hawaii. The W.M. Keck Observatory is operated
as a scientific partnership among the California Institute of
Technology, the University of California and NASA and was made possible
by the generous financial support of the W.M. Keck Foundation.}

\end{abstract}


\section{Introduction}

Exploring the era when the first stars formed by locating high redshift
sources with demonstrably young cosmic ages represents the next
outstanding challenge for observational cosmology (Mather \& Stockman
2000). Although luminous quasars (Zheng et al 2000, Fan et al 2000,
2001) and star-forming galaxies (Dey et al 1998, Weymann et al 1998,
Spinrad et al 1998, Hu et al 1999) have been located beyond $z\simeq$5,
to be detected these must be spectacularly luminous and rare examples
drawn from a largely unknown underlying population (for an excellent
review of attempts to find very distant galaxies, see Stern \& Spinrad
1999).

Gravitational magnification by foreground clusters of galaxies, whose
mass distributions are constrained by arcs and multiple images of known
redshift, has already provided new information on the abundance of
faint background objects (Kneib et al 1996). Particularly high
magnifications ($\simeq\times$40) are expected in the {\it critical
regions} which can be located precisely in well-understood clusters for
sources occupying specific redshift ranges, e.g. $2<z<\,$7. Although
the volumes probed in this way are smaller than those addressed in
panoramic narrow band surveys (Hu et al 1998, Malhotra et al 2001),
intrinsically much fainter and most likely more representative sources
are sampled. If the surface density of such sources is sufficient, this
may be a promising route for securing the first glimpse of young cosmic
sources beyond $z\simeq$5.

Accordingly, we have begun a blind spectroscopic survey of the
appropriate critical lines of several well-constrained lensing clusters
with Hubble Space Telescope images (Santos et al 2001).  Briefly, our
strategy involves undertaking long-slit scans of regions 7 $\times$ 120
arcsec in extent with the Keck I Low Resolution Imaging Spectrograph
(LRIS, Oke et al 1995), using gratings that offer a spectral resolution
of $\simeq$4 \AA\ in the OH forest and $\simeq$6 \AA\ in the blue. The
typical wavelength range covered is $\lambda\lambda$3500-9350
\AA\ offering the potential of seeing lensed Ly $\alpha$ sources in the
important range 2$<z_s<$7.  With a 1.0 arcsec slit, the dwell at each
location is normally 2 $\times$ 1000sec.

In the course of surveying the cluster Abell 2218 ($z$=0.18) on 23
April 2001 we encountered a strong emission line at $\lambda$7989
\AA\ close to the central cD (Figures 1 and 2).  Astrometry associates
this emission with a faint, marginally-resolved, source in the Early
Release WFPC2 F814W image (labelled {\it a} in Figure 1) with
$I_{814}$=25.9 $\pm$ 0.2. Inspection of Kneib et al's (1996) mass model
suggests that a second image with $I_{814}\simeq$26.0 $\pm$ 0.3, 6
arcsec away ({\it b} in Figure 1), represents a counter-image of the
same highly magnified $z>$5 source.

On May 21, 2001 we used the Keck II Echelle Spectrograph and Imager
(ESI, Scheinis et al 2000) at a higher spectral resolution ($\simeq$
1.25 \AA\ ) with a 0.75 arcsec slit aligned to include both images
(see inset panel in Figure 1). With 2 $\times$ 2000 sec exposures,
strong emission was confirmed from {\it both} images (Figure 3a).  The
spectra are identical (to within the signal/noise) confirming the
lensing hypothesis.  Importantly, the magnitude difference in the emission
lines ($\Delta m_{line}\simeq$0.2$\pm$0.1) is comparable to that in the
$I_{814}$ photometry. The combined flux-calibrated spectrum (Figure 3b)
reveals a single emission line with an asymmetric (P Cygni-like)
profile suggestive of gas outflow.

The location and separation of the images was already suggestive of
lensing of a high redshift source consistent with
emission arising from Ly$\alpha$ at $z$=5.576 (corresponding to the
peak in the combined spectrum at $\lambda$7996 \AA\ \footnote{This
redshift is presumably a slight overestimate by an unknown amount given
the likelihood of self-absorption.}). Were the emission to arise from
$H\alpha$, the images have to be a physically associated pair just
behind the cluster and the absence of other emission would be puzzling
given the extensive LRIS wavelength coverage. The most plausible
alternative to Ly$\alpha$ for a lone emission line would be [O\,II] at
$z$=1.14.  This can be eliminated not only by lensing arguments (c.f.
the location of the critical lines and image configurations expected in
Figure 1), but also by the fact the [O\,II] 3726, 3728 \AA\ doublet
would be readily resolved at the spectral resolution of ESI.

%\centerline{\psfig{file=fig1.eps,width=7cm}}
\centerline{\psfig{file=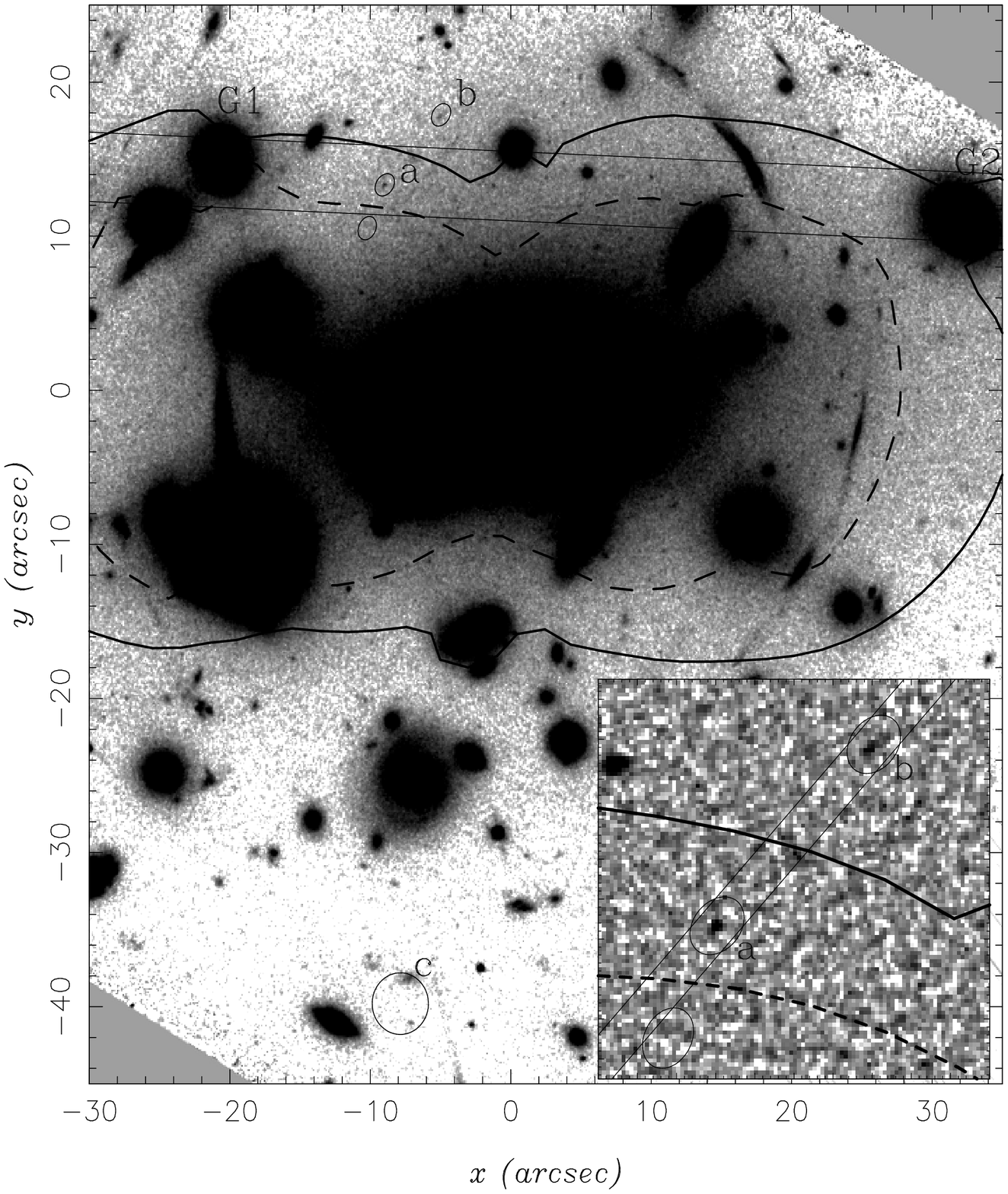,width=14cm}}
\noindent{\bf Figure 1:}\quad {\em Hubble Space Telescope F814W image
of Abell 2218 (z=0.18) with the location of the LRIS longslit scanning
region marked. {\rm a} and {\rm b} represent the lensed pair at
z=5.576; the inset panel (10 $\times$ 10 arcsec) illustrates the
secondary spectroscopic configuration adopted with ESI. Curves refer to
critical lines of infinite magnification for a source at $z$=1.14
(dashed) and 5.576 (solid) in the context of Kneib et al's (1996) mass
model. For a source at z=1.14, the counter-image of {\rm a} would lie
just below the appropriate critical line (as indicated by the small
circle) and is not seen. The large circle {\rm c} refers to the region
where a much fainter ($I\sim$29) third image is expected for a source
at z=5.576.}
\medskip

%\centerline{\psfig{file=fig2.eps,width=8cm}}
\centerline{\psfig{file=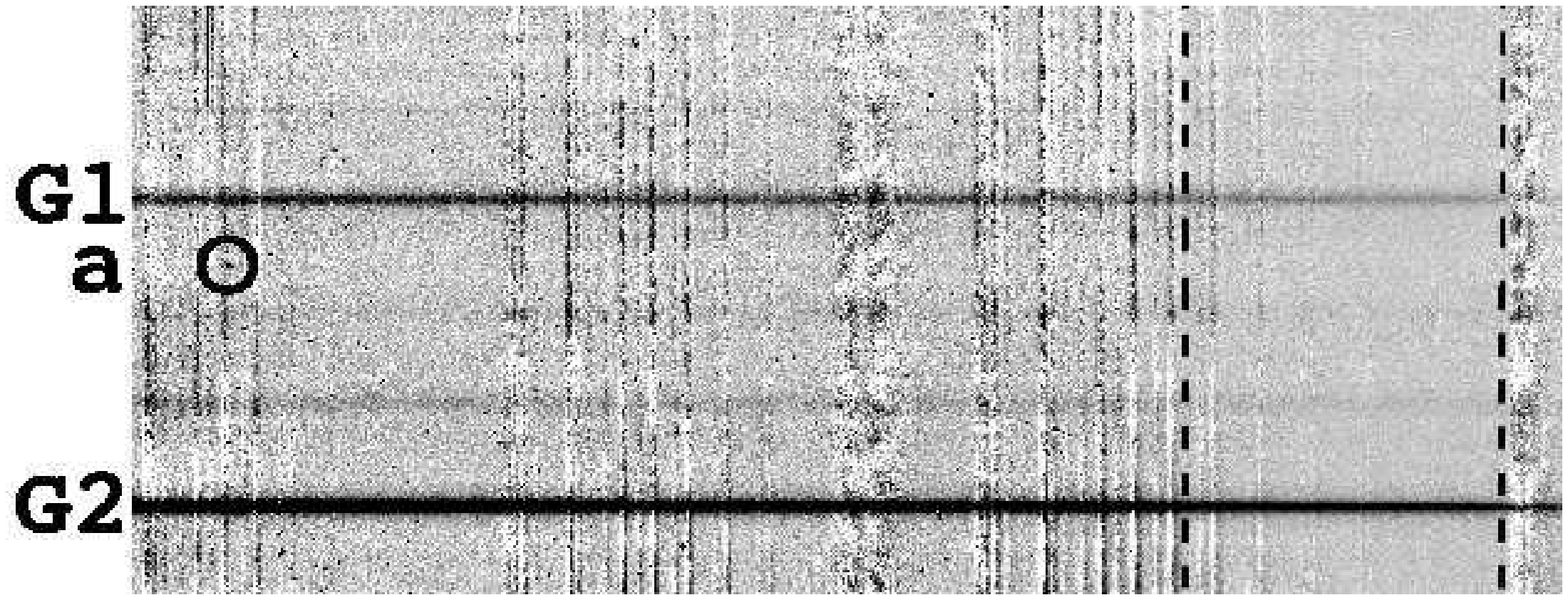,width=16cm}}
\noindent{\bf Figure 2:}\quad {\em The discovery of an emission line
source close to the cD in the rich cluster Abell 2218.  Keck I LRIS-R
spectral image of a region 100 arcsec in extent covering
$\lambda\lambda$6700-9350 \AA\ with the emission line attributed to
object {\rm a} at $\lambda$7989 \AA\ marked.  The dashed lines at
longer wavelengths refer to the wavelength range used to deduce a
statistical upper limit on a stellar continuum from the source (see
text). The spectra of fiducial cluster galaxies {\rm G1} and {\rm G2}
labelled in Figure 1 are marked.}
\medskip

\centerline{\hbox{
\psfig{file=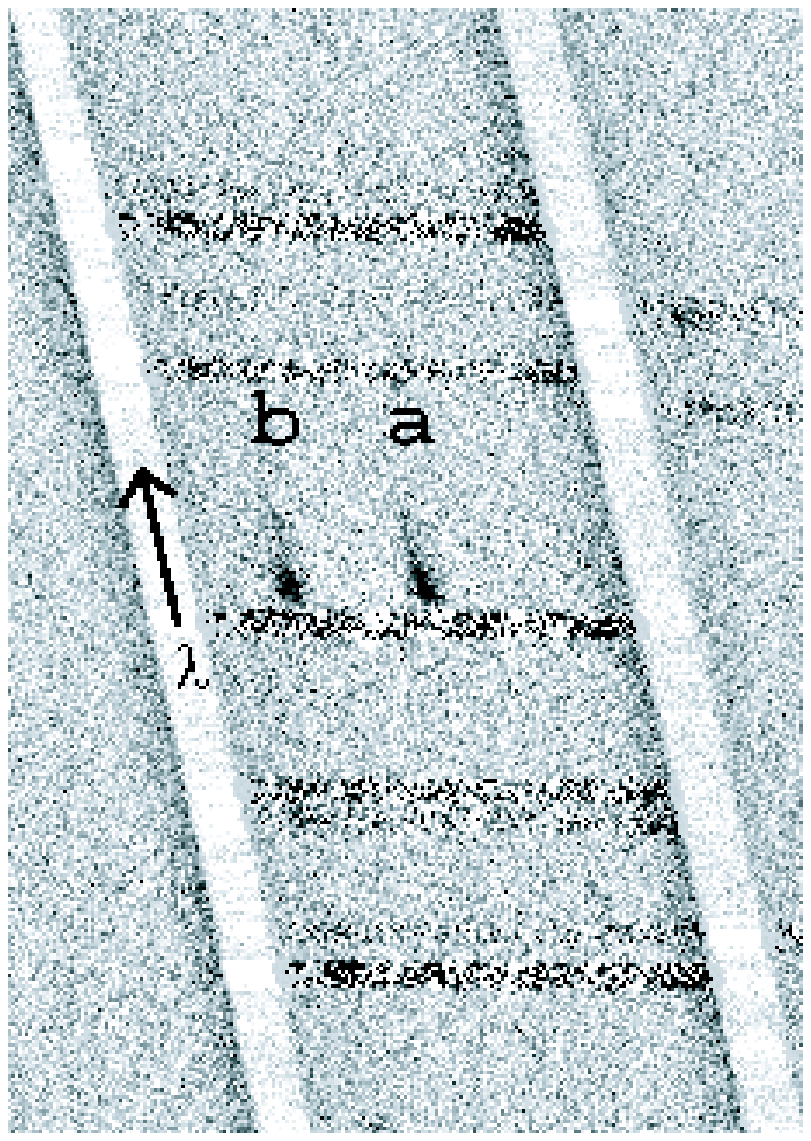,width=5.8cm}
\psfig{file=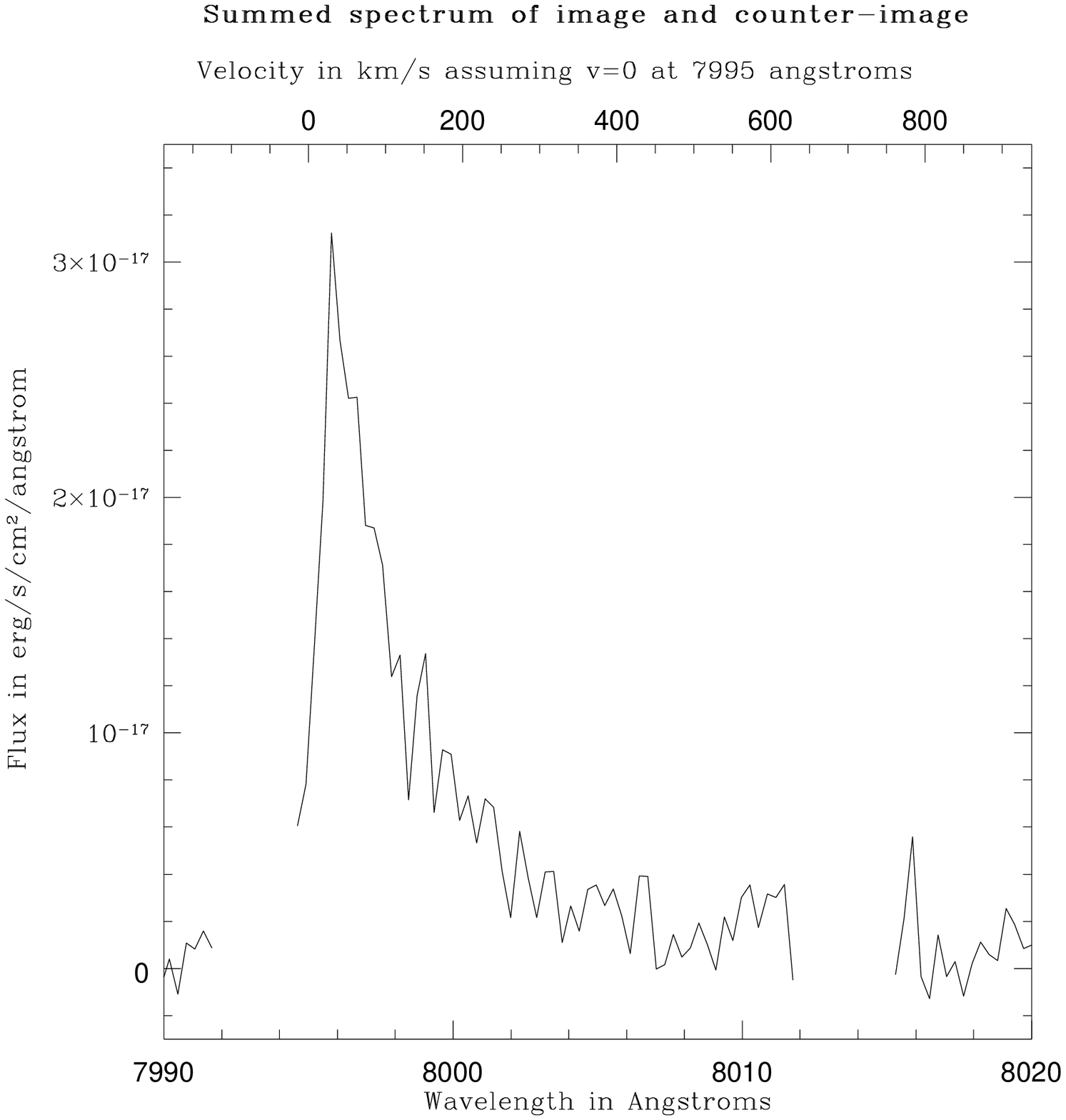,width=8.3cm} }}

\noindent{\bf Figure 3:}\quad {\em Confirmation of strong emission
in the pair of images marked in Figure 1 using the Keck II
Echelle Spectrograph and Imager. (a) 2-D sky subtracted spectral
image using the 20 arcsec slit. (b) Flux-calibrated spectrum of
the region around Lyman alpha emission combined from both images
revealing a P-Cygni like profile extending redward by
$\simeq$200 km sec$^{-1}$ in the rest-frame. The redshift
corresponding to the peak emission is $z$=5.576.}

\section{Source Properties}

The remarkable features of the $z$=5.576 source are its faintness
(particularly considering the high magnification afforded
by its proximity to the critical line), its small angular size in
the HST image, and the apparent absence of any stellar continuum
in both the LRIS and ESI spectra. 

The magnification of the two images in Figure 1b can be determined
from the Abell 2218 mass model (Kneib et al 1996) which has been 
extensively tested via spectroscopy of 18 arclets by Ebbels et al
(1999). In this model, the magnifications for $a$ and $b$ are,
respectively 3.8 mag ($\times$33.1) and 3.7 mag ($\times$30.2)
implying a (unlensed) source magnitude of $I_{814}\simeq$29.7.
Inspection of the dithered WFPC2 image indicates that image $a$ is
marginally resolved along the shear direction (i.e. towards the other
image). The appropriate half-light scales are 0.23 $\times$ $<$0.15
arcsec. Allowing for the HST resolution and the linear magnification at
this point in the cluster's gravitational field implies a physical
diameter of less than 150 $h_{65}^{-1}$ pc \footnote{We assume a
cosmological model with $\Omega_M$=0.3 and $\Omega_{\Lambda}$=0.7
througout.}.

The lensing model also offers insight into the crucial question of
whether we are witnessing magnification of an isolated object or a
star-forming component (e.g. a HII region) embedded in a more extended
source close to a caustic. The mass model indicates that the source
that produces the pair lies 1.2 kpc from the caustic. Thus any
comparable emitting region (containing line or continuum flux) within
this distance would also be highly magnified and possibly detected.
Together with the remarkably small physical size, this suggests the
source is a truly isolated system and not, for example, a star forming
sub-component of a larger luminous system (c.f. Franx et al 1997,
Trager et al 1997).

A substantial component of the broad-band $I$-band flux arises
from the line emission suggesting that the stellar continuum is 
unusually faint. If the F814W flux were produced by a single emission line
at $\lambda$7989, the flux density in the line would be $F_a(HST)$=1.2
$\pm$ 0.2 10$^{-16}$ ergs cm$^{-2}$ sec$^{-1}$. This is only 70\%
higher than the mean inferred from the ESI spectra, corrected for extinction:
$F_a(ESI)$= 6.8$\pm$0.7 10$^{-17}$ ergs cm$^{-2}$ sec$^{-1}$.
The ESI line flux is consistent, within uncertainties of absolute
calibration, with that inferred for $a$ in the LRIS data:\,
$F_a(LRIS)$=5.6 $\pm$ 0.5 10$^{-17}$ ergs cm$^{-2}$ sec$^{-1}$.

Limits on any stellar continuum flux can be explored further in
the LRIS wavelength region $\lambda\lambda$9020--9297 \AA\ which is
relatively free from OH contamination (Figure 2). Including the 
noise across the LRIS slit at this location we deduce a 3$\sigma$ 
upper limit to the continuum flux of 3.\,$10^{-20}$ ergs cm$^{-2}$ 
sec$^{-1}$ \AA\ $^{-1}$. Assuming a flat spectrum longward of Ly$\alpha$, 
this upper limit integrated over the F814W bandpass would also yield 
a signal comparable to the emission line flux.

Limited near-infrared data is available for Abell 2218 from
commissioning data taken with the INGRID infrared camera on the
4.2m William Herschel Telescope (supplied by courtesy of Ian
Smail). Image $a$ remains undetected to limits of $J$=22.5 and
$K$=21.5 (5$\sigma$ for a point source). At respective rest-frame
wavelengths $\lambda\simeq$1600 and 3350 \AA\ , neither filter 
is likely to be contaminated by a strong emission lines. These
non-detections give further constraints on the continuum flux,
viz. $F_a(J)<3.9\,10^{-19}$ ergs cm$^{-2}$ sec$^{-1}$ \AA\ $^{-1}$
and $F_a(K)< 9.8\, 10^{-20}$ ergs cm$^{-2}$ sec$^{-1}$ \AA\
$^{-1}$.

We summarize the properties of the source detected in Abell 2218
in Table~1.  Although our observed line flux is comparable to those in
sources seen at lower redshift in narrow band searches (Hu et al 1998),
when lensing is taken into account the true source flux is much
fainter. 

\begin{deluxetable}{lrrr}
\tablecaption{Unlensed Source Fluxes}
\tablehead{\colhead{Dataset} & \colhead{Total} 
         & \colhead{continuum\tablenotemark{a}} &\colhead{Ly-$\alpha$}}
\startdata
& \multicolumn{1}{c}{mag}
& \multicolumn{1}{c}{erg cm$^{-2}$ s$^{-1}$ \AA$^{-1}$}
& \multicolumn{1}{c}{erg cm$^{-2}$ s$^{-1}$} \\
HST/WFPC2 F814W &  29.7  &$<2.4\ 10^{-20}$&$< 3.6\pm0.6\ 10^{-18}$ \\
LRIS            &        &$<6\ 10^{-21}$  &  $1.7\pm0.2\ 10^{-18}$ \\
ESI             &        &                &  $2.1\pm0.2\ 10^{-18}$ \\
WHT J           &$>26.3$ &$<8\ 10^{-20}$  &                        \\
WHT K           &$>25.3$ &$<2\ 10^{-20}$  &                        \\
\enddata
\medskip
\quad\tablenotemark{a}\quad{3-$\sigma$ upper limits on the continuum flux, 
per unit wavelength in the rest frame. }
\end{deluxetable}

%The rest-frame equivalent width ($W_{\alpha}>$ 300 \AA\ ),
%whilst uncertain, appears to be much higher than those encountered in
%normal star-forming systems (c.f. Charlot \& Fall 1993).

\section{First Light?}

We now address the interesting question of whether the source lensed by
Abell 2218 is being observed at a special time in its history, perhaps
consistent with its first generation of stars.  Although the Ly$\alpha$
line is an unreliable guide to the ongoing star formation rate because
of self-absorption, scattering and dust extinction difficulties, will
argue that uncertainties arising from this diagnostic most likely
strengthen our conclusions.

Adopting the relationship 1 $M_{\odot}$ yr$^{-1}$ = 1.5 10$^{42}$ ergs
sec$^{-1}$ in Ly$\alpha$ (Ferland \& Osterbrock 1985, Kennicutt 1998,
Osterbrock 1989, ) and including a magnification of 33 with a
100\% escape fraction and zero extinction, we infer a current star
formation rate (SFR) of 0.5 $M_{\odot}$ yr$^{-1}$. We consider
this a lower limit given the conservative assumptions above. Although 
our physical scale of $<$150pc is comparable to that resolved for 30 
Doradus in the Large Magellanic Cloud (Scowen et al 1998), the SFR 
is over an order of magnitude larger than the integrated value
for energetic giant H II regions contained within nearby star-forming 
galaxies (McKee \& Williams 1997). Consistent with its isolated nature, 
the source appears to be a very powerful extragalactic HII 
region with a luminosity $L_{\alpha}\simeq$10$^{42}$ ergs sec$^{-1}$ 
(c.f. Melnick et al 2000).

At $z$=5.576, in our adopted cosmology, the cosmic age is only 1 Gyr.
We ran the Starburst99 code (Leitherer et al 1999) for a metal-poor
($Z$=10$^{-3}Z_{\odot}$) system  with a constant SFR of 0.5 $M_{\odot}$
yr$^{-1}$ in order to explore at what age a detectable stellar
continuum would emerge in the LRIS spectral window
($\lambda_{UV}=\lambda_{rest}$=1370--1415 \AA\ ). Ignoring dust
extinction, this provides a tighter constraint than the same
calculation applied to the $J$ and $K$ band limits at their longer rest
wavelengths. For our adopted upper limit of $F_{UV}<3.\,$10$^{-20}$
ergs cm$^{-2}$ \AA\ $^{-1}$ (see \S2), the appropriate unlensed
continuum luminosity, $L_{UV} < 2.\,10^{39}$ ergs sec$^{-1}$
\AA\ $^{-1}$, would be exceeded at the observed SFR in less than 2 Myr
suggesting the object could be remarkably young with a stellar mass
$\sim10^6 M_{\odot}$.

If the SFR were higher in the past, or if the Ly$\alpha$
emission were subject to upward corrections due to
self-absorption, the implied age for the continuum flux limit
would be even shorter. Although we cannot yet provide any observable
constraints on dust extinction, given the Ly$\alpha$ line is more
likely to be suppressed than the adjacent continuum, this would 
also imply that we have {\it overestimated} the age and implied 
stellar mass. 

\section{Discussion}

Hierarchical models of structure formation predict a high density of
systems undergoing their first era of star formation at $z\simeq$6
(Haiman \& Spaans 1999). Our critical line survey (Santos et al 2001)
will provide new constraints on their abundance and redshift 
distribution out to $z\simeq$7. In particular, the example discussed 
here could not have been detected without the lensing boost afforded by 
Abell 2218. Its unlensed equivalent would not have been reliably 
detected even in the Hubble Deep Field.

The most interesting suggestion arising from our study is the possible
young age inferred from our upper limit on the stellar continuum in the
context of the star formation rate deduced from the Ly$\alpha$ flux.
While there are many uncertainties in this deduction, we argue they
work in the sense of strengthening the conclusion.  If our upper age
limit is correct, very deep infrared imaging would be needed to
reliably probe the spectral energy distribution of this source longward
of 1 $\mu$m, i.e. in the rest-frame optical. Depending on the star
formation history, lensed 2$\mu$m fluxes of 50 nJy ($K\simeq$25) are
expected.  An unlensed analog would have a flux density of only 1 nJy
and would clearly be challenging even for NGST.

HII regions of stellar mass of order 10$^6$ $M_{\odot}$ with star
formation rates of $\simeq$ 1\,$M_{\odot}$ yr$^{-1}$ can be found at
lower redshifts. The significance of the system in Abell 2218 lies in
the fact that an isolated, possibly young, low mass system has been
located close to the redshift at which many now believe re-ionization
may be occurring (Djorgovski et al 2001, Becker et al 2001). Just as
with those constraints which sample a few (possibly atypical)
sightlines to a distant quasar, so the stellar history of further
examples of our star-forming source, located with the aid of strong
lensing, will provide an early census of such systems beyond
$z\simeq$5.

%\newpage

\acknowledgements We thank Fred Chaffee, Carlos Frenk and Hyron Spinrad
for their encouragement and acknowledge useful discussions with Andrew
Benson, Claus Leitherer, Dan Stern, Tommaso Treu and Pieter van
Dokkum.  Richard Massey and Pieter van Dokkum are thanked for
assistance in reducing the Keck observations. Ian Smail is thanked for
access to the WHT infrared observations. Faint object spectroscopy at
the Keck observatory is made possible through the dedicated efforts of
Joe Miller, Mike Bolte and colleagues at UC Santa Cruz (for ESI), and
Judy Cohen, Bev Oke, Chuck Steidel and colleagues at Caltech (for
LRIS).

\end{document}